\documentclass [12pt]{article}

\begin{document}

\title{\textbf{ Effective bosonic hamiltonian for excitons
: \\a too naive concept}}
\author{M. Combescot and O.
Betbeder-Matibet\\GPS, Universit\'e  Denis
Diderot and Universit\'e Pierre et Marie Curie,
CNRS,\\Tour 23, 2 place Jussieu, 75251 Paris
C\'edex 05, France}
\date{}
\maketitle

\begin{abstract}
Excitons, being made of two fermions, may
appear from far as bosons. Their
close-to-boson character is however quite
tricky to handle properly. Using our commutation technique
especially designed to deal with interacting
close-to-boson particles, we here calculate the
\emph{exact} expansion in Coulomb interaction of the
exciton-exciton correlations, and show that a naive
effective bosonic
hamiltonian for excitons cannot produce
these X-X correlations correctly. 
\end{abstract}

\vspace{2cm}


\newpage

Up to now, problems dealing with interacting    
particles have in common the fact that the
hamiltonian can be separated into
$H=H_0+V$.The eigenstates of the
non-interacting part $H_0$ of this hamiltonian
being usually known exactly, they can serve as
an orthogonal basis for the system, so that
expansions in the interaction $V$ are simply
obtained by using the identity 
\begin{eqnarray}
\frac{1}{a-H}
& = & \frac{1}{a-H_0}+\frac{1}{a-H}V\frac{1}{a-H_0}\nonumber
 \\ & = & \frac{1}{a-H_0}+\frac{1}{a-H_0}V\frac{1}{a-H_0}
 +\frac{1}{a-H_0}V\frac{1}{a-H_0}V\frac{1}{a-H_0}+\ldots
\end{eqnarray}
(with $a$ being usually $(\omega + i\eta)$), and by inserting the
closure relation for
$H_0$ eigenstates on both sides of the $V$ operators
\nolinebreak  $^{(1)}$.

The problem of interacting excitons is much more tricky
because it is not at all of that type. The excitons are
indeed the exact \emph{one} electron-hole pair eigenstates
of the semiconductor hamiltonian $H$ and thus can serve as
an orthogonal basis for any one-pair state of the
semiconductor. However when there is one pair only in the
system, there is no X-X interaction ! The problem
starts with two pairs as there is no way $^{(2)}$ to extract from
the semiconductor hamiltonian $H$ an $H_0$ part for which
the product of two excitons would be the \emph{exact}
eigenstates and so could serve as an orthogonal basis for
two-pair states.

The fundamental difficulty with interacting excitons comes
from the fact that they are composite particles made of two
fermions, so that they are not "clean" particles. There are
\emph{a priori} two ways to couple two electrons and two
holes to make two excitons. However these two ways are hard
to cope with the fact that electrons and holes are
indiscernable particles. Moreover, as electrons and holes
are fermions, it is physically obvious that two excitons
must "feel" each other not only through Coulomb
interaction between their carriers but also through Pauli
exclusion between their indiscernable components,
\emph{even in the absence of any Coulomb interaction} :
This Pauli exclusion is in fact the extremely subtle part
of the interacting exciton problem.

Attempts have been made, using various bosonisation
procedures $^{(3,4)}$, to replace the exact semiconductor
hamiltonian
$H$ by an effective hamiltonian $^{(5)}$,
\pagebreak
\begin{eqnarray}
H_\mathrm{eff} & = &
H_\mathrm{eff}^0+V_\mathrm{eff}\nonumber
\\
&
=
&
\sum_{i}E_i\bar{B}_i^\dag\bar{B}_i+\frac{1}{2}\sum_{mnij}\mathcal{E}_{mnij}\bar{B}_{m}^\dag\bar{B}_{n}^\dag\bar{B_i}\bar{B_j},
\end{eqnarray}
in which the excitons are assumed to be
bosonic particles,
$\left[\bar{B}_i,\bar{B}_j^\dag\right]=\delta_{ij}$,
provided that the underlying fermionic character of their
components is included in an appropriate exciton
interaction $\mathcal{E}_{mnij}$. We have recently shown $^{(2)}$
that the proposed $V_\mathrm{eff}$ cannot be correct :
First, it is not an hermitian operator 
as the given  $^{(6)}$ $\mathcal{E}_{mnij}$ is not equal to 
$\mathcal{E}_{ijmn}^\ast$. Even if this mistake is
corrected, the bosonisation procedures actually dress the
exact Coulomb interaction operator by exchange processes
which originate from Pauli exclusion between carriers, so
that by construction they miss purely Pauli terms, i.\ e.\ terms
without
$e^2/r$ factors. We have shown $^{(2)}$ that these Pauli
 terms are precisely those necessary to restore the
hermiticity of the effective exciton hamiltonian quoted 
up to now.

In the present work, we show that, in addition to the
existence of these conceptually new purely Pauli terms,
Pauli exclusion, when handled properly, dresses the Coulomb
scatterings in such a subtle way that we do not see how
a dressed exciton-exciton Coulomb interaction can
produce the exchange Coulomb terms of the exciton-exciton
correlations correctly. In other words, in the low density limit, it
would be nice to replace the exact semiconductor hamiltonian by an
effective hamiltonian for boson-excitons ; but, unfortunately, the
form quoted in Eq (2) is too naive $^{(5)}$ to be satisfactory.

Our work relies on the fact that, although $H$ cannot be written
as $H_0+V$, with $V$ describing interactions
between excitons, it is nevertheless possible to generate
an exact expansion in Coulomb interaction, in its spirit
similar to Eq (1), owing to our commutation technique. This
technique is briefly summarized in section 1 for excitons
without spin degrees of freedom. While easy to include $^{(7)}$,
these spin parameters make the notations quite heavy so
that they tend to hide the subtle physics induced by Pauli
exclusion.

In section 2, we show how our commutation technique allows
to calculate the X-X correlations exactly at any order in
Coulomb interactions. From a trivial algebra, we see that
exchange processes induced by Pauli exclusion enter the
correlation terms in a far from obvious way. From the form of
these correlation terms, we see no way to find an exciton-exciton
scattering which corresponds to Coulomb interactions dressed by
exchange processes and which can be valid beyond
first order in Coulomb interaction !

\section{Survey of the commutation technique}

Our commutation technique $^{(2,7)}$ allows to calculate any
quantity dealing with interacting excitons in terms of two
parameters $\xi_{mnij}^\mathrm{dir}$ and $\lambda_{mnij}$.

The first one comes from Coulomb interaction. It appears
through 
\begin{equation}\left[V_i^\dag,B_j^\dag\right]=\sum_{m,n}
\xi_{mnij}^\mathrm{dir}B_m^\dag B_n^\dag,
\end{equation}
where the operator $V_i^\dag$ describes $^{(8)}$ the Coulomb
interaction between the $i$ exciton and the rest of the
system. It is defined by
\begin{equation}\left[H,B_i^\dag\right]=E_iB_i^\dag+V_i^\dag
,\end{equation} 
where $H$ is the exact semiconductor hamiltonian, and
$B_i^\dag$ the exact (one) exciton creation operator,
$HB_i^\dag \mid v>=E_iB_i
^\dag \mid v>$, $E_i$ being the
$i$ exciton energy, and $\mid v>$ the vacuum state for
electron-hole pairs.

The second parameter is linked to the close-to-boson
character of the excitons induced by Pauli exclusion. It
appears through
\begin{equation}\left[D_{ni},B_j^\dag\right]=2\sum_m
\lambda_{mnij}B_m^\dag ,\end{equation}
where $D_{ni}$ is the deviation-from-boson operator defined
by $\left[B_n,B_i^\dag\right]=\delta_{ni}-D_{ni}$. We note
that, if the excitons were exact bosons, $D_{ni}\equiv 0$,
so that $\lambda_{mnij}$ would be zero. This
$\lambda_{mnij}$ parameter also appears when one couples
the electrons and holes of two excitons in a different way.
We then get $^{(2)}$
\begin{equation}
 B_i^\dag B_j^\dag
= - \sum_{m,n}\lambda_{mnij}B_m^\dag B_n^\dag. 
\end{equation}
The fact that $B_i^\dag B_j^\dag
\mid v >$ is not a well
defined state as it contains a piece of any other two-exciton
state, makes the exciton problem quite tricky. In particular, these
$B_i^\dag B_j^\dag
\mid v >$ states do not form an orthogonal basis as
\begin{equation}
\frac{1}{2!}< v \mid B_mB_nB_i^\dag B_j^\dag \mid v
>=\hat{\delta}_{mnij}-\lambda_{mnij},
\end{equation}
which is easy to deduce from Eq (5).
This matrix element thus differs from zero even if
 $\hat{\delta}_{mnij}=(\delta_{mi}\delta_{nj}+\delta_{mj}\delta_{ni}
)/2$ is zero, i.\ e.\ even if $(m,n) \neq (i,j)$.

A clear physical understanding of these two crucial
parameters for interacting excitons,
$\xi_{mnij}^\mathrm{dir}$ and $\lambda_{mnij}$, is easy to
get from their expressions in $\mathbf{r}$ space. The
(direct) Coulomb parameter is equal to $^{(2)}$
\begin{eqnarray}
\xi_{mnij}^\mathrm{dir}=\frac{1}{2}
\int de_1\,de_2\,dh_1\,dh_2\,\phi_m^\ast
(e_1,h_1)\phi_n^\ast (e_2,h_2)\left[ V_{e_1 e_2}+V_{h_1
h_2}-V_{e_1 h_2}-V_{e_2 h_1}\right]\nonumber \\ \times
\phi_i(e_1,h_1)\phi_j(e_2,h_2)  +  (m\longleftrightarrow n),
\end{eqnarray}
where $V_{eh} = e^2/\mid \mathbf{r}_e - \mathbf{r}_h \mid$
while $\phi_i(e,h)$ is the total wave function of the $i$
exciton with its electron in 
$\mathbf{r}_e$ and its hole in
$\mathbf{r}_h$. $\xi_{mnij}^\mathrm{dir}$ thus
corresponds to \emph{all Coulomb processes between (m,n)
and (i,j) excitons} when these excitons are built on the
\emph{same} pairs $(e_1,h_1)$ and $(e_2,h_2)$. The
$\lambda_{mnij}$ parameter reads in \textbf{r} space,
\begin{equation}
\lambda_{mnij}=\frac{1}{2}\int de_1\,de_2\,dh_1\,dh_2\,
\phi_m^\ast (e_1,h_1) \phi_n^\ast (e_2,h_2)
\phi_i (e_1,h_2) \phi_j(e_2,h_1)+(m\longleftrightarrow n),
\end{equation}
so that it simply describes the possible exchange of the
two electrons (or holes) when building the $(m,n)$ and
$(i,j)$ excitons.

In many physical quantities also appear the combinations,
\begin{equation}
\xi_{mnij}^\mathrm{right}=\sum_{r,s}\xi_{mnrs}^\mathrm{dir}
\lambda_{rsij}, \hspace{.5in}
\xi_{mnij}^\mathrm{left}=\sum_{r,s}\lambda_{mnrs}\xi_{rsij}^\mathrm{dir},
\end{equation} 
which are linked by $^{(2)}$
\begin{equation}
(E_m+E_n-E_i-E_j)\lambda_{mnij}=\xi_{mnij}^\mathrm{left}-\xi_{mnij}^\mathrm{right}.
\end{equation} 
Their half sum reads in \textbf{r} space,
\begin{eqnarray}
 \xi_{mnij}^\mathrm{exch} =
\frac{1}{2}\left(\xi_{mnij}^\mathrm{right}+\xi_{mnij}^\mathrm{left}\right)
=  \frac{1}{2}\int
de_1\,de_2\,dh_1\,dh_2\,\phi_m^\ast(e_1,h_1)\phi_n^\ast(e_2,h_2)
\hspace{.5in}
\nonumber
\\  \times 
\left[V_{e_1e_2}+V_{h_1h_2}-(V_{e_1h_1}+V_{e_2h_2}+V_{e_1h_2}+V_{e_2h_1})/2\right]
\phi_i(e_1,h_2)\phi_j(e_2,h_1)+(m\longleftrightarrow
n),
\end{eqnarray}
so that $\xi_{mnij}^\mathrm{exch}$ contains \emph{
all possible Coulomb processes} \emph{ between the two
electrons and the two holes of the two excitons}, when
these $(m,n)$ and
$(i,j)$ excitons are built with \emph{different}
electron-hole pairs. $\xi_{mnij}^\mathrm{right}$ reads as
$\xi_{mnij}^\mathrm{exch}$ with the e-h contribution being
$\left( V_{e_1h_2}+V_{e_2h_1}\right)$, while for
$\xi_{mnij}^\mathrm{left}$ this e-h contribution is $\left(
V_{e_1h_1}+V_{e_2h_2}\right)$.

Let us end this survey by noting that, as obvious from Eqs
(8,9,10,12), all the $\xi$ parameters are homogeneous to an
energy, while the $\lambda$ parameter is dimensionless.

\section{Exciton-exciton correlations}

In plenty of problems, the hamiltonian $H$
appears through matrix elements of $1/(a-H)$ where
$a$ is usually $(\omega + i \eta)$. As the excitons are the
exact one-pair eigenstates of $H$, the matrix elements of
$1/(a-H)$ between one-exciton states are simply
\begin{equation}
<v\mid B_j \frac{1}{a-H} B_i^\dag \mid v >=\frac{1}{a-E_i}
\delta_{ij}.
\end{equation}
The difficulties arrive with two-pair states because
$HB_i^\dag B_j^\dag \mid v >\neq (E_i+E_j)B_i^\dag B_j^\dag
\mid v >$, so that
\begin{equation}
G_{mnij}(a)=\frac{1}{2!}<v\mid B_mB_n \frac{1}{a-H}B_i^\dag
B_j^\dag \mid v>
\end{equation}
is not equal to
\begin{eqnarray} 
G_{mnij}^0 (a) & = &
\frac{\hat{\delta}_{mnij}}{a-E_i-E_j}=\hat{\delta}_{mnij}g_{ij}(a)
 \\  & = &  \frac{1}{2!} <v\mid \bar{B}_m \bar{B}_n
\frac{1}{a-H_{\mathrm{eff}}^0} \bar{B}_i^\dag
\bar{B}_j^\dag \mid v>,
\end{eqnarray}
as it would be for excitons interacting neither by Coulomb
interaction nor by Pauli exclusion. The matrix elements of
$1/(a-H)$ between two-exciton states have additional
terms we are now going to calculate.

A clean way to have the Coulomb interactions between
excitons appearing is via the operators $V_i^\dag$ defined
in Eq (4). From this equation, we get \mbox{ $B_i^\dag
(a-H-E_i)=(a-H)B_i^\dag +V_i^\dag$}, so that we do have
\begin{equation}
\frac{1}{a-H}B_i^\dag=B_i^\dag
\frac{1}{a-H-E_i}+\frac{1}{a-H}V_i^\dag \frac{1}{a-H-E_i}
\hspace{.1in} .
\end{equation} 
Eq (17) is, in its spirit, very similar to Eq (1) which
allows to derive correlation effects for usual problems
dealing with interactions. This Eq (17) is in fact the key
equation for problems dealing with correlation
effects between excitons : For two excitons, it leads to
\begin{equation}
\frac{1}{a-H}B_i^\dag B_j^\dag \mid v>=\frac{1}{a-E_i-E_j}
\left(B_i^\dag+\frac{1}{a-H} V_i^\dag \right) B_j^\dag \mid
v> ,
\end{equation}
with a similar equation for $<v\mid
B_mB_n\left(1/(a-H)\right)$. From them and Eqs (3,7), we
can obtain the integral equations verified by the matrix
elements of $1/(a-H)$ in the two-exciton subspace. They read
\begin{eqnarray}
G_{mnij}(a)&=&G_{mnij}^0 (a)-\sum_{pqrs} G_{mnpq}^0
(a)X_{pqrs}^\mathrm{right} (a)G_{rsij}^0
(a)+\sum_{pqrs}G_{mnpq}(a)\xi_{pqrs}^\mathrm{dir}
G_{rsij}^0 (a) \hspace{1cm} \\ &=&G_{mnij}^0
(a)-\sum_{pqrs} G_{mnpq}^0 (a)X_{pqrs}^\mathrm{left}
(a)G_{rsij}^0 (a)+\sum_{pqrs}G_{mnpq}^0
(a)\xi_{pqrs}^\mathrm{dir} G_{rsij}(a),  
\end{eqnarray}
where we have set
\begin{equation}
X_{mnij}^\mathrm{right}(a)=(a-E_m-E_n)\lambda_{mnij},\hspace{.7in}
X_{mnij}^\mathrm{left}(a)=\lambda_{mnij}(a-E_i-E_j).
\end{equation}
The $X$'s appear as the vertices one has to associate to
Pauli exclusion between two excitons. Due to Eq (11), the
$X$'s and the $\xi$'s are linked by
\begin{equation}
X_{mnij}^\mathrm{left}(a)-X_{mnij}^\mathrm{right}(a)=\xi_{mnij}^\mathrm{left}-\xi_{mnij}^\mathrm{right}.
\end{equation}
These Pauli vertices are however rather peculiar since,
depending on $a$, they depend on the particular problem of
interest. We can note that the Coulomb interactions dressed
by exchange processes, $\xi_{mnij}^\mathrm{right}$ and
$\xi_{mnij}^\mathrm{left}$ introduced previously, are just
a sequence of a direct Coulomb scattering and a right or
left Pauli scattering before or after it :
\begin{equation}
\xi_{mnij}^\mathrm{right}=\sum_{pqrs}\xi_{mnpq}^\mathrm{dir}G_{pqrs}^0
(a)X_{rsij}^\mathrm{right}(a) \hspace{.6in}
\xi_{mnij}^\mathrm{left}=\sum_{pqrs}X_{mnpq}^\mathrm{left}(a)G_{pqrs}^0
(a)\xi_{rsij}^\mathrm{dir}.
\end{equation}
Let us however stress that, while the $X$'s depend on $a$,
neither $\xi^\mathrm{left}$ nor $\xi^\mathrm{right}$
depend on it.

Before going further, let us note that the
integral equations (18-19) for $G_{mnij}(a)$ are indeed rather
nice as they look like Dyson equations $^{(9)}$, $G=G^0 +G^0 VG$,
except for the additional Pauli term which is $G^0 X G^0$ and not
$G^0 XG$. This "little" change makes all the subtle
effects induced by Pauli exclusion between excitons.

As for $\xi_{mnij}^\mathrm{exch}$, it will be convenient to
introduce the Pauli scattering
\begin{equation}
X_{mnij}(a)=\frac{1}{2}\left[
X_{mnij}^\mathrm{right}(a)+X_{mnij}^\mathrm{left}(a)
\right] =\left( a-\frac{E_i+E_j+E_m+E_n}{2} \right)
\lambda_{mnij}.
\end{equation}
By inserting Eq (19) into Eq (20) and \emph{vice versa}, and by
using Eqs (12,23,24), the integral equations verified by
$G_{mnij}(a)$ take a form more symmetrical with respect to
the left and right exchange processes :
\begin{eqnarray}
G_{mnij}(a)=G_{mnij}^0 (a)+g_{mn}(a)
\left[
\xi_{mnij}^\mathrm{dir}-\xi_{mnij}^\mathrm{exch}-X_{mnij}(a) \right]
g_{ij}(a) \nonumber 
\\ +g_{mn}(a)\left(
\sum_{pqrs}\xi_{mnpq}^\mathrm{dir}G_{pqrs}(a)\xi_{rsij}^\mathrm{dir}\right)
g_{ij}(a),
\end{eqnarray}
with $g_{ij}(a)$ defined in Eq (15). This leads to 
\begin{equation}
G_{mnij}(a)=G_{mnij}^0 (a)+g_{mn}(a)\left[
\xi_{mnij}^\mathrm{dir}-\xi_{mnij}^\mathrm{exch}+\xi_{mnij}^\mathrm{corr}(a)-X_{mnij}(a)\right]
g_{ij}(a),
\end{equation}
where $\xi_{mnij}^\mathrm{corr}(a)$ corresponds to all
direct and exchange correlation processes of the $(i,j)$
excitons into the $(m,n)$ states, with two or more Coulomb
interactions. From Eqs (23-25), we find that
$\xi_{mnij}^\mathrm{corr}(a)$ verifies the integral equation
\begin{equation}
\xi_{mnij}^\mathrm{corr}(a)=\Xi_{mnij}^{(2)}
(a)+\Xi_{mnij}^{(3)}(a)+\sum_{pqrs}\xi_{mnpq}^\mathrm{dir}g_{pq}(a)\xi_{pqrs}^\mathrm{corr}(a)g_{rs}(a)\xi_{rsij}^\mathrm{dir},
\end{equation}
where $\Xi_{mnij}^{(2)}(a)$ contains all direct and
exchange second order contributions in $\xi$,
namely,
\begin{equation}
\Xi_{mnij}^{(2)}(a)=\sum_{pq}\left[
\xi_{mnpq}^\mathrm{dir}g_{pq}(a)\xi_{pqij}^\mathrm{dir}-
\frac{1}{2}\xi_{mnpq}^\mathrm{right}g_{pq}(a)\xi_{pqij}^\mathrm{dir}
-\frac{1}{2}\xi_{mnpq}^\mathrm{dir}g_{pq}(a)\xi_{pqij}^\mathrm{left}\right],      
\end{equation}
while $\Xi_{mnij}^{(3)}(a)$ contains all
third order contributions,
\begin{equation}
\Xi_{mnij}^{(3)}(a)=\sum_{pqrs}\xi_{mnpq}^\mathrm{dir}g_{pq}(a)\left[
\xi_{pqrs}^\mathrm{dir}-\xi_{pqrs}^\mathrm{exch}\right]
g_{rs}(a)\xi_{rsij}^\mathrm{dir}.
\end{equation}

Let us examine the equations (26-29) giving the matrix elements of
$1/(a-H)$ in the two-exciton subspace. From the first order
contribution in Coulomb interaction, i.\ e.\  in $\xi$, we
could think that Pauli exclusion dresses the bare Coulomb
interaction by transforming $\xi^\mathrm{dir}$ into
$\xi^\mathrm{dir}-\xi^\mathrm{exch}$. A rapid glance at the
higher order processes immediately shows that the way the
exchange processes enter the correlation terms is much more subtle,
and that the linear combination
$\xi^\mathrm{dir}-\xi^\mathrm{exch}$ appearing at first
order in $\xi$, does not remain unchanged at higher order :
While odd terms in Coulomb interaction do depend on
$\xi^\mathrm{exch}$, this symmetrical combination of
$\xi^\mathrm{right}$ and $\xi^\mathrm{left}$ does not
appear in the even terms, as they contain the exchange
Coulomb processes in different positions in the sequence of
direct Coulomb interactions. Moreover, we see that the
exchange contributions appear \emph{once only}, in the
"middle" of a set of $\xi^\mathrm{dir}$.

As a major consequence, this shows that there is little hope to
identify an effective boson-exciton interaction resulting
from direct and exchange Coulomb processes which could
produce the expansion of the $1/(a-H)$ matrix
elements exactly : Indeed if an effective hamiltonian like the one
of Eq (2) were to exist, the expansion would be : \pagebreak
\begin{eqnarray} 
G_{mnij}^{\mathrm{eff}} & = & \frac{1}{2!}<v \mid
\bar{B}_m\bar{B}_n \frac{1}{a-H_\mathrm{eff}}\bar{B}_i^\dag
\bar{B}_j^\dag \mid
v>\nonumber
\\ & = &
G_{mnij}^0(a)+g_{mn}(a)\ \mathcal{E}_{mnij}\ g_{ij}(a)\nonumber
\\ & + &
\sum_{pq}g_{mn}(a)\ \mathcal{E}_{mnpq}\ g_{pq}(a)\
\mathcal{E}_{pqij}\  g_{ij}(a)+\cdots
\end{eqnarray}
with the \emph{same} scattering $\mathcal{E}$ between the $g$
factors.

Let us end by stressing that the scattering of two excitons
from $(i,j)$ to $(m,n)$ states also contains a purely Pauli
contribution $X_{mnij}(a)$, i.\ e.\ a contribution which
does not contain any $\xi$ Coulomb interaction at all (see
Eqs (24-26)). This is conceptually quite new and very
specific to close-to-boson particles, which feel each other
through Pauli exclusion between their components, quite
independently from any Coulomb scattering.

\section{Conclusion}
By using our commutation technique, we have derived the
\emph{exact} Coulomb expansion of the $1/(a-H)$ matrix
elements in the two-exciton subspace. From the obtained
expression of the exciton-exciton correlations, we show that
the exchange processes appear in such a subtle way in the
sequence of direct Coulomb scatterings, that there is
no hope for a naive effective Coulomb
interaction between boson-excitons to take into account
their composite nature properly.

\newpage
\hbox to \hsize{\hfill REFERENCES \hfill}
\vspace{1cm}
(1) C. COHEN-TANNOUDJI, B. DIU, F. LALOE, \emph{MŽcanique
Quantique} (Hermann, Paris, 1973). 

(2) M. COMBESCOT, O. BETBEDER-MATIBET, Cond. mat. 0106346 (18 June
2001) ; Europhysics Letters (accepted). 

(3) T. USUI, Prog. Theor. Phys. \underline{23}, 787 (1957).

(4) For a review see for instance : A. KLEIN, E.R. MARSHALEK, Rev.
Mod. Phys. \underline{63}, 375 (1991).

(5) Higher order terms like $BBBB^{\dag} B^{\dag} B^{\dag}$ are
sometimes mentioned in this effective hamiltonian. It is however
clear that they cannot help for the point raised here, as we
consider matrix elements in the \emph{two}-exciton subspace, so that
all these higher order terms would give zero in this subspace.

(6) See for example Eqs (5.7-9) in H. HAUG, S. SCHMITT-RINK, Prog.\ 
Quant.\  Elect.\ \underline{9}, 3 (1984). 

(7) O. BETBEDER-MATIBET, M. COMBESCOT, submitted to Eur. Phys. J. B.

(8) M. COMBESCOT, R. COMBESCOT, Phys.\ Rev.\ Lett.\ \underline{61},
117 (1988).

(9) A. FETTER, J. WALECKA, \emph{Quantum Theory of
Many-Particle Systems} (McGraw-Hill, New York, 1971).

\end{document}